\documentclass{article}
\def\err#1#2{\stackrel{\scriptstyle +#1}{\scriptstyle -#2}}

\begin{document}
\centerline{Date: \today     \hfill{\bf UCD-2000-8} }
\centerline{                 \hfill{\bf MAD-00-1163} }
\vspace*{.5in}
\begin{center}
{\large Atomic Parity Violation, Leptoquarks, and Contact Interactions} \\
\vskip 0.7cm
V. Barger$^a$ and Kingman Cheung$^b$ \\
$^a$ {\it Department of Physics, University of Wisconsin, 
1150 University Ave., Madison, WI 53706}\\
$^b$ {\it Department of Physics, University of California, Davis, 
CA 95616 USA} \\
\end{center}

\begin{abstract}
The recent measurement of atomic parity violation in cesium atoms
shows a $2.3\sigma$ deviation from the standard model prediction.  We
show that such an effect can be explained by four-fermion contact
interactions with specific chiralities or by scalar leptoquarks which
couple to the left-handed quarks.  For a coupling of electromagnetic
strength, the leptoquark mass is inferred to be 1.1 to 1.3 TeV. We
also show that these solutions are consistent with all other
low-energy and high-energy neutral-current data.
\end{abstract}

\section{}

Parity violation in the standard model (SM) results from exchanges of
weak gauge bosons.  In electron-hadron neutral-current processes parity
violation is due to the vector axial-vector 
interaction terms in the Lagrangian.  These interactions have been tested to
a high accuracy in atomic parity violation (APV) measurements.
A very recent measurement 
in cesium (Cs) atoms has been reported \cite{apv} by measuring a parity-odd
transition between the $6S$ and $7S$ energy levels of the Cs atoms.  
The measurement is stated in
terms of the weak charge $Q_W$, which parameterizes the parity violating
Hamiltonian.

The new measurement of the atomic parity violation in cesium 
atoms is \cite{apv}
\begin{equation}
\label{first}
Q_W ( ^{133}_{55} {\rm Cs} ) = -72.06 \pm 0.28 ({\rm expt}) \pm 0.34 
({\rm theo})\;.
\end{equation}
This result represents 
a substantial improvement over the previously reported value
\cite{oldapv},
because of a more precise calculation of the atomic wavefunctions \cite{wave}.
Compared to the standard model prediction $Q_W^{\rm SM}= -73.09 \pm 0.03$ 
\cite{sm-value}, the deviation $\Delta Q_W$ is
\begin{equation}
\label{data}
\Delta Q_W \equiv Q_W({\rm Cs}) - Q_W^{\rm SM}({\rm Cs}) 
= 1.03 \pm 0.44\;,
\end{equation}
which is $2.3\sigma$ away from the SM prediction.

In this Letter, we propose leptoquark solutions to this APV
measurement and also solutions with four-fermion contact
interactions.  We find that the weak-isospin-doublet 
leptoquark ${\cal S}^R_{1/2}$, which
couples to the right-handed electron and left-handed $u$ and $d$
quarks, and the weak-isospin-triplet 
leptoquark $\vec{\cal S}_1^L$, which couples to left-handed
electron and left-handed $u,d$ quarks,
can explain the measurement with the coupling-to-mass
ratio $\lambda/M \sim 0.29$ and 0.24 TeV$^{-1}$,
respectively, where
$\lambda$ is the coupling and $M$ is the
leptoquark mass.  For a coupling of electromagnetic strength the
leptoquark masses are 1.1 to 1.3 TeV.  We verify that these
leptoquark explanations are comfortably consistent with
all existing experimental constraints.  We also find that
contact interactions with $\eta_{RL}^{eu} = \eta_{RL}^{ed}=-0.043$
TeV$^{-2}$ and others can alternatively explain the APV measurement and are
consistent with a global fit to data on $\ell\ell qq$ interactions.

Another possible explanation for the APV measurement is extra $Z$ bosons
\cite{extra-z}, which can come from a number of grand-unified theories.
Previous work on constraining new physics using the atomic parity violation
measurements can be found in Ref. \cite{prev}.

\section{}
The parity-violating part of the Lagrangian describing electron-nucleon
scattering is given by 
\begin{equation}
{\cal L}^{eq} = \frac{G_F}{\sqrt{2}} \sum_{q=u,d} \left\{
C_{1q}( \bar e \gamma^\mu \gamma^5 e ) (\bar q \gamma_\mu q)
+C_{2q}( \bar e \gamma_\mu e ) (\bar q \gamma^\mu \gamma^5 q) \right \}
\end{equation}
where in the SM the coefficients $C_{1q}$ and $C_{2q}$ at tree level are given
by
\[
C_{1q}^{\rm SM} = - T_{3q} + 2 Q_q \sin^2\theta_{\rm w}\;, \qquad
C_{2q}^{\rm SM} = - T_{3q} (1 - 4 \sin^2\theta_{\rm w})\;.
\]
Here $T_{3q}$ is the third component of the isospin of the quark $q$ and 
$\theta_{\rm w}$ is the weak mixing angle.  In terms of the $C_{1q}$, the
weak charge $Q_W$ for Cs is $Q_W = -376 C_{1u} - 422 C_{1d}$.  Since we are
interested in the deviation of $Q_W$ from its SM value, we write
\begin{equation}
\Delta Q_W ({\rm Cs}) = -376 \Delta C_{1u} - 422 \Delta C_{1d} \;.
\end{equation}

A convenient form \cite{ours} for four-fermion $eeqq$ contact interactions is 
\cite{quigg}
\begin{equation}
{\cal L}_\Lambda = \sum_{q=u,d} \left \{
 \eta_{LL} \overline{e_L} \gamma_\mu e_L \overline{q_L} \gamma^\mu q_L 
+\eta_{LR} \overline{e_L} \gamma_\mu e_L \overline{q_R} \gamma^\mu q_R 
+\eta_{RL} \overline{e_R} \gamma_\mu e_R \overline{q_L} \gamma^\mu q_L 
+\eta_{RR} \overline{e_R} \gamma_\mu e_R \overline{q_R} \gamma^\mu q_R 
\right \} \;,
\end{equation}
where $\eta_{\alpha\beta} = 4\pi \epsilon/(\Lambda^{eq}_{\alpha\beta})^2$
and $\epsilon=\pm1$.  The contact interaction contributions to the 
$\Delta C_{1q}$'s are
\begin{equation}
\Delta C_{1q} = \frac{1}{2\sqrt{2} G_F} \left [
-\eta_{LL}^{eq} + \eta_{RR}^{eq} - \eta_{LR}^{eq} + \eta_{RL}^{eq} \right ]\;,
\end{equation}
and the corresponding contributions to $\Delta Q_W$ are
\begin{equation}
\label{th}
\Delta Q_W = ( -11.4\; {\rm TeV}^{2} ) \left[
-\eta_{LL}^{eu} + \eta_{RR}^{eu} - \eta_{LR}^{eu} + \eta_{RL}^{eu} \right ]
+ 
( -12.8\; {\rm TeV}^{2} ) \left[
-\eta_{LL}^{ed} + \eta_{RR}^{ed} - \eta_{LR}^{ed} + \eta_{RL}^{ed} \right ]
\;.
\end{equation}

\begin{table}[t]
\caption{\small \label{table1}
The values of $\eta_{\alpha\beta}^{eu,ed}$ required to fit the $\Delta Q_W$
data of Eq. (\ref{data}).  We assume one nonzero $\eta$ at a time.
}
\medskip
\centering
\begin{tabular}{|cc||cc|}
\hline
$\eta$ & fitted value (TeV$^{-2}$) & $\eta$ & fitted value (TeV$^{-2}$) \\
\hline
\hline
$\eta_{LL}^{eu}$ & $ 0.090$ & $\eta_{LL}^{ed}$ & $0.081 $\\
$\eta_{RR}^{eu}$ & $-0.090$ & $\eta_{RR}^{ed}$ & $-0.081$ \\
$\eta_{LR}^{eu}$ & $ 0.090$ & $\eta_{LR}^{ed}$ & $0.081 $\\
$\eta_{RL}^{eu}$ & $-0.090$ & $\eta_{RL}^{ed}$ & $-0.081$ \\
\hline
\end{tabular}
\end{table}

In order to explain the data in Eq. (\ref{data}) using contact interactions,
we can apply Eq. (\ref{th}) with nonzero $\eta$'s.  However, from 
Eq. (\ref{th}) we see that there could be cancellations among the 
$\eta$-terms.
When we assume one nonzero $\eta$ at a time, the values required to fit the
APV data are tabulated in Table \ref{table1}.  The value of $\Lambda$ 
corresponding to $\eta=0.090 (0.081)$ TeV$^{-2}$ is $11.8 (12.5)$ TeV.  If 
we further assume a SU(2)$_L$ symmetry, then $\eta_{RL}^{eu}$ equals
$\eta_{RL}^{ed}$ and the value to fit the APV data is 
\begin{equation}
\label{contact}
\eta_{RL}^{eu} = \eta_{RL}^{ed} = -0.043\; {\rm TeV}^{-2}\;,
\end{equation}
which corresponds to a $\Lambda \sim 17$ TeV.  
Equation (\ref{contact}) is relevant to one of the leptoquark solutions that 
we present in the next section.

The next question to ask is whether the above solutions are in conflict with
other existing data.  To answer this, we performed an analysis \cite{ours} 
of the neutral-current lepton-quark contact interactions using 
a global set of $\ell\ell qq$ data, which includes
(i) the neutral-current (NC) deep-inelastic scattering at HERA,
(ii) Drell-Yan production at the Tevatron,
(iii) the hadronic production cross sections at LEPII,
(iv) the parity violation measurements in $e$-(D, Be, C) scattering at
SLAC, Mainz, and Bates, 
(v) the $\nu$-nucleon scattering measurements by CCFR and NuTeV, and
(vi) the lepton-hadron universality of weak charged-currents.
This is an update of the analysis in Ref. \cite{ours} with new 
data from LEPII, finalized and published data from H1 and ZEUS \cite{nc}, and 
including D\O\ data on Drell-Yan production \cite{dy-d0}.
The 95\% C.L. one-sided limits 
on $\eta$'s and the corresponding limits on $\Lambda$
are given in Table \ref{table2}.  In obtaining these limits, we do not include
the data on atomic parity violation, which is the new physics data that we 
want to describe in this paper.  

In Table \ref{table2}, the most tightly constrained are $\eta_{LL}^{eu}$ and
$\eta_{LL}^{ed}$, mainly due to the constraint of lepton-hadron universality
of weak charged currents.  In general, the constraints on $eu$ parameters 
are stronger than those on $ed$ parameters, because of Drell-Yan production,
in which the $u\bar u$-initial-state channel is considerably more important 
than the $d\bar d$-initial-state channel.
{}From Table \ref{table2} the 95\% C.L. 
one-sided limits on $\eta_{RL}$ are 0.30 TeV$^{-2}$ and $-0.64$ TeV$^{-2}$ for
$\epsilon=+$ and $\epsilon=-$, respectively.  Thus, the
fit to the APV data in Eq. (\ref{contact}) lies comfortably within the limits
and so are the solutions with $\eta_{LR}$ and $\eta_{RR}$. 
On the other hand, the solution using $\eta_{LL}^{eu}$ is ruled out
while the solution using $\eta_{LL}^{ed}$ is marginal.

\begin{table}
\caption{\small \label{table2}
The 95\% C.L. one-sided limits on $\eta_{\alpha\beta}^{eq},\; 
\alpha,\beta=L,R, q=u,d$.  The ``$+$'' and ``$-$'' signs correspond to the
$\epsilon$ in the definition of $\eta$'s.  The corresponding limits on
$\Lambda_{\alpha\beta}^{eq}$ are also shown.  The SU(2)$_L$ implied 
relation $\eta_{RL}^{eu}=\eta_{RL}^{ed}$ is included. }
\medskip
\centering
\begin{tabular}{|cc|cc|cc|}
\hline
$\eta$ & fitted value (TeV$^{-2}$) 
& \multicolumn{2}{c|}{$\eta_{95}$ (TeV$^{-2}$)} & 
\multicolumn{2}{c|}{$\Lambda$ (TeV)} \\
\hline
\hline
      &    & {$+$} &  {$-$} & 
             {$+$} &  {$-$} \\
$\eta_{LL}^{eu}$ & $-0.057 \pm 0.030$ & 0.034 & $-0.11$ & 19.4 & 10.8 \\
$\eta_{LR}^{eu}$ & $-0.024 \pm 0.15$ & 0.28 & $-0.32$ & 6.6 & 6.3 \\
$\eta_{RL}^{eu}=\eta_{RL}^{ed}$ & $-0.38 \err{0.20}{0.17}$ &
 0.30 & $-0.64$ & 6.4 & 4.4 \\
$\eta_{RR}^{eu}$ & $-0.23\err{0.15}{0.14}$ & 0.20 & $-0.46$ & 7.9 & 5.2 \\
\hline
$\eta_{LL}^{ed}$ & $0.059\pm{0.033}$ & 0.11 & $-0.037$ & 10.5 & 18.6 \\
$\eta_{LR}^{ed}$ & $-0.048\err{0.33}{0.31}$ & 0.62 & $-0.60$ & 4.5 & 4.6\\
$\eta_{RR}^{ed}$ & $0.32\err{0.26}{0.30}$ & 0.73 & $-0.61$ & 4.1 & 4.5\\
\hline
$\eta_{LL}^{eu}=\eta_{LL}^{ed}/2$ 
 & $0.058\pm{0.034}$ & 0.11 & $-0.040$ & 10.5 & 17.8 \\
\hline
\end{tabular}
\end{table}

\section{}
The Lagrangians representing the interactions of the $F=0$ and $F=-2$ 
($F$ is the fermion number)
scalar leptoquarks are \cite{buch,rizzo}
\begin{eqnarray}
\label{9}
{\cal L}_{F=0} &=& \lambda_L \overline{\ell_L} u_R {\cal S}_{1/2}^L
+ \lambda_R \overline{q_L} e_R (i \tau_2 {\cal S}^{R*}_{1/2} )
+ \tilde{\lambda}_L \overline{\ell_L} d_R \tilde{{\cal S}}_{1/2}^L + h.c. \;,\\
\label{10}
{\cal L}_{F=-2} &=& g_L \overline{q_L^{(c)}} i \tau_2 \ell_L {\cal S}_0^L
+ g_R \overline{u_R^{(c)}} e_R {\cal S}_0^R
+ \tilde{g}_R \overline{d_R^{(c)}} e_R \tilde{{\cal S}}_0^R
+ g_{3L}\overline{q_L^{(c)}} i \tau_2 \vec{\tau} \ell_L \cdot \vec{\cal S}_1^L
+ h.c.
\end{eqnarray}
where $q_L,\ell_L$ denote the left-handed quark and lepton doublets, 
$u_R,d_R,e_R$ denote the right-handed up quark, down quark, and 
electron singlet, and $q_L^{(c)}, u_R^{(c)}, d_R^{(c)}$ denote the 
charge-conjugated fields.
The subscript on leptoquark fields denotes the weak-isospin of the leptoquark,
while the superscript ($L,R$) denotes the handedness of the lepton that
the leptoquark couples to.  
The components of the $F=0$ leptoquark fields are
\begin{equation}
{\cal S}_{1/2}^{L,R} = \left ( \begin{array}{c}
        {S_{1/2}^{L,R} }^{(-2/3)} \\
        {S_{1/2}^{L,R} }^{(-5/3)}  \end{array} \right ) \;, \;\;\;\;\;
\tilde{{\cal S}}_{1/2}^L = \left( \begin{array}{c}
       \tilde{S}_{1/2}^{L(1/3)} \\
     - \tilde{S}_{1/2}^{L(-2/3)} \end{array} \right ) \;,
\end{equation} 
where the electric charge of the component fields is given in the 
parentheses, and the corresponding hypercharges are $Y({\cal S}_{1/2}^L)=
Y({\cal S}_{1/2}^R)=-7/3$ and $Y(\tilde{{\cal S}}_{1/2}^L)=-1/3$.
The $F=-2$ leptoquarks ${\cal S}_0^L, {\cal S}_0^R, \tilde{{\cal S}}_0^R$
are isospin singlets with hypercharges $2/3, 2/3, 8/3$, respectively, while
${\cal S}_1^L$ is a triplet with hypercharge $2/3$:
\begin{equation}
{\cal S}_1^L = \left( \begin{array}{l}
               { S_1^L }^{(4/3)} \\
               { S_1^L }^{(1/3)} \\
               { S_1^L }^{(-2/3)} \end{array} \right ) \;.
\end{equation}
The SU(2)$_L\times$ U(1)$_Y$ symmetry is assumed in the Lagrangians
of Eqs. (\ref{9}) and (\ref{10}).

We have verified that the contributions of leptoquarks ${\cal S}^L_{1/2}$, 
$\tilde{\cal S}^L_{1/2}$, ${\cal S}_0^R$, and $\tilde{\cal S}_0^R$, that
couple to the right-handed quarks,
only give a negative $\Delta Q_W$, which cannot explain
the measurement in Eq. (\ref{first}). 
The only viable choices are the leptoquarks ${\cal S}^R_{1/2}$,
${\cal S}_0^L$, and $\vec{\cal S}_1^L$ that couple to the
left-handed quarks.  Let us first examine the contribution from the
$F=0$ leptoquark ${\cal S}^R_{1/2}$.  The effective interaction of 
electron-quark scattering via this leptoquark is
\begin{equation}
{\cal L} = - \frac{\lambda_{R}^2}{M^2_{ {\cal S}^R_{1/2}}} \left(
\overline{d_L} e_R \overline{e_R}d_L + \overline{u_L} e_R \overline{e_R}u_L
\right ) \;,
\end{equation}
where we have assumed $M^2_{ {\cal S}_{1/2}^R} \gg s,|t|,|u|$ and 
the overall negative sign is due to the ordering of the fermion 
fields relative to the $\gamma,Z$ diagrams.  
After a Fierz transformation, the above amplitude
can be transformed to \begin{equation}
{\cal L} = - \frac{\lambda_{R}^2}{2 M^2_{{\cal S}^R_{1/2}}} \left(
 \overline{e_R} \gamma^\mu e_R  \overline{d_L} \gamma_\mu d_L 
+ \overline{e_R} \gamma^\mu e_R \overline{u_L} \gamma_\mu u_L  \right )\;.
\end{equation}
Comparing with the contact interaction terms, we can relate the above
equation to $\eta_{RL}$ as
\begin{equation}
\eta_{RL}^{eu} = \eta_{RL}^{ed} = - \frac{\lambda_{R}^2}
{2 M^2_{{\cal S}^R_{1/2}}} \;.
\end{equation}
Using the result on contact terms in Eq. (\ref{contact}) and the above
equation, we obtain the value for $\lambda_{R}/M_{{\cal S}^R_{1/2}}$ to be
\begin{equation}
\label{final}
\frac{\lambda_{R}}{M_{{\cal S}^R_{1/2}}} = 0.29 \; {\rm TeV}^{-1} \;.
\end{equation}
This result cannot specifically indicate the value for the mass or the
coupling of the leptoquark, 
because the APV is a low-energy atomic process that only probes the
$\lambda_{R}/M_{{\cal S}^R_{1/2}}$ ratio.  

Similarly, the effective interaction of electron-quark scattering involving
${\cal S}^L_0$ is
\begin{equation}
{\cal L} = \frac{g_L^2}{2 M^2_{ {\cal S}_0^L}} 
\overline{e_L} \gamma^\mu e_L \overline{u_L} \gamma_\mu u_L \;.
\end{equation}
Therefore, the contribution from ${\cal S}_0^L$, in terms of contact 
interaction, is
\begin{equation}
\eta_{LL}^{eu} = \frac{g_L^2}{2 M^2_{ {\cal S}_0^L} }\;.
\end{equation}
Matching with the results in Table \ref{table1} the coupling-to-mass ratio
of the leptoquark is given by
\begin{equation}
\label{final2}
\frac{g_L}{M_{ {\cal S}_0^L} } = 0.43\; {\rm TeV}^{-1}\;.
\end{equation}
However, this leptoquark ${\cal S}_0^L$ contributes $\eta_{LL}^{eu}
=0.09$ TeV$^{-2}$ and that is ruled out by the limit in Table \ref{table2}.

The interaction of the $F=-2$ leptoquark $\vec{\cal S}_1^L$ is given by
\begin{equation}
{\cal L} = g_{3L} \left \{
-\left
(\overline{u_L^{(c)}}e_L + \overline{d_L^{(c)}} \nu_L \right) 
\, {\cal S}_1^{L(1/3)}
- \sqrt{2}\; \overline{d_L^{(c)}} e_L \; {\cal S}_1^{L(4/3)}
+ \sqrt{2}\; 
\overline{u_L^{(c)}} \nu_L \; {\cal S}_1^{L(-2/3)} + h.c. \right \}
\;.
\end{equation}
The effective interaction of electron-quark scattering involving 
$\vec{\cal S}_1^L$ is
\begin{equation}
{\cal L} = \frac{ g_{3L}^2}{2 M^2_{ {\cal S}_1^L} }\; 
\overline{e_L} \gamma^\mu e_L \; \overline{u_L} \gamma_\mu u_L +
\frac{ g_{3L}^2}{ M^2_{ {\cal S}_1^L} }\; 
\overline{e_L} \gamma^\mu e_L \; \overline{d_L} \gamma_\mu d_L  \;.
\end{equation}
Therefore, the contributions from $\vec{\cal S}_1^L$, in terms of
contact interaction, are
\begin{equation}
\eta_{LL}^{eu} = \frac{ \eta_{LL}^{ed}}{2} 
= \frac{g_{3L}^2}{2 M^2_{ {\cal S}_1^L} }\;.
\end{equation}
Fitting to $\Delta Q_W$ using Eq. (\ref{th}), we obtain the coupling-to-mass
ratio of $\vec{\cal S}_1^L$ to be
\begin{equation}
\label{3L}
\frac{g_{3L}}{M_{ {\cal S}_1^L}} = 0.24 \; {\rm TeV}^{-1} \;,
\end{equation} 
which gives $\eta_{LL}^{eu}=0.028\;{\rm TeV}^{-2}$ and 
$\eta_{LL}^{ed}=0.056\;{\rm TeV}^{-2}$.  We recalculate the limit from the
global set of neutral-current $\ell\ell qq$ data for the 
case of nonzero $\eta_{LL}^{eu}$ and $\eta_{LL}^{ed}$ with 
$\eta_{LL}^{eu} = \eta_{LL}^{ed}/2$.  We obtain the 95\% C.L. one-sided
limits on $\eta_{LL}^{eu}=\eta_{LL}^{ed}/2$ as $0.11\; {\rm TeV}^{-2}$
and $-0.04\;{\rm TeV}^{-2}$ for $\epsilon=+$ and $\epsilon=-$, respectively
(this result is listed in the last row of Table \ref{table2}.)
Therefore, this leptoquark $\vec{\cal S}_1^L$ solution is also 
consistent with all other data.

As discussed above, there are two leptoquark solutions that are consistent 
with the limits in Table \ref{table2}.  The one with the $F=0$ leptoquark 
${\cal S}_{1/2}^R$ requires the coupling-to-mass ratio equal 0.29 TeV$^{-1}$.
With a coupling strength about the same as $e=0.31$, the inferred
leptoquark mass is about 1.1 TeV for ${\cal S}_{1/2}^R$.  
Similarly, the coupling-to-mass ratio for the $F=-2$ leptoquark 
$\vec{\cal S}_1^L$ is required to be 0.24 TeV$^{-1}$, which corresponds to
 a mass of 1.3 TeV.

\section{}
In the following we discuss the above leptoquarks that describe
the APV measurement in the context of the
limits from various collider experiments.  

The model-independent search
for the first generation scalar leptoquark at the Tevatron by CDF and D\O\ puts
a lower bound of 242 GeV on the mass of the leptoquark \cite{cdf-d0}.
The direct search for the first generation scalar 
leptoquark at HERA, on the other hand,
depends on the coupling constants and the type of the leptoquark.
ZEUS \cite{zeus} excluded the first generation scalar leptoquark 
(fermion number 
$F=0$) with electromagnetic coupling strength up to a mass of 280 GeV 
while H1 \cite{h1} excluded a mass up to 275 GeV in
$e^+ p$ collisions.  In the most recent searches in $e^- p$ collisions,
ZEUS excluded $F=-2$ scalar leptoquarks up to about 290 GeV mass \cite{zeus}.
In general, $e^\pm p$ colliders can search for leptoquarks up to mass 
almost equal to the center-of-mass energy of the machine.  

The LEP collaborations performed both direct searches for leptoquarks
and indirect searches for virtual effects of leptoquarks in fermion-pair
production.  OPAL \cite{opal} searched for real leptoquarks in pair
production and excluded scalar leptoquarks up to about 88 GeV;
DELPHI \cite{delphi} searched for leptoquarks in single production and excluded
scalar leptoquarks up to about 161 GeV.
Various LEP Collaborations \cite{lep} analyzed fermion-pair production and were
able to rule out some leptoquark coupling and mass ranges, which depend 
sensitively on the leptoquark type and couplings.  The best mass 
limit is around
300 GeV for electromagnetic coupling strength.  The virtual effects
in fermion-pair production have already been included in our global
analysis presented in Sec. 2.
If we take $\lambda_{R}$ and $g_{3L}$ of
electromagnetic strength, the leptoquark masses are inferred to be
1.1 and 1.3 TeV, respectively, as already noted  above.
Therefore, the solutions in Eqs. (\ref{final}) and (\ref{3L}) lie 
comfortably with
both the direct search limits and the virtual effects in neutral-current
$\ell\ell qq$ data.  

An important low-energy constraint to leptoquarks or contact
interactions is lepton-hadron universality of weak
charged-currents (CC), which we have already included in the global
analysis in Sec. 2.  Since it is particularly important to leptoquark
interactions, we would like to explain it briefly.  Because of the
SU(2)$_L$ symmetry, the $\eta_{LL}^{eu}$ and $\eta_{LL}^{ed}$ are
related to the CC contact interaction $\eta_{CC}
\overline{\nu_L}\gamma_\mu e_L \overline{d_L}\gamma^\mu u_L$ by
$\eta_{LL}^{ed} - \eta_{LL}^{eu} = \eta_{CC}$.  Thus, the NC contact
interactions and leptoquarks are subject to the constraint on
$\eta_{CC}$.  The leptoquarks that are constrained by this $\eta_{CC}$
are ${\cal S}_0^L$ and ${\cal S}_1^L$, which couple to the left-handed
leptons and quarks.  The CC contact
interaction $\eta_{CC} \overline{\nu_L}\gamma_\mu e_L
\overline{d_L}\gamma^\mu u_L$ could upset two important
experimental constraints: (i) lepton-hadron universality in weak CC,
and (ii) $e$-$\mu$ universality in pion decay, of which the former
gives a stronger constraint.  Using the values for the CKM matrix
elements the constraint on $\eta_{CC}$ is $2 \eta_{CC} = (0.102 \pm
0.073)\; {\rm TeV}^{-2}$ \cite{ours}.  It is mainly because of this
constraint that the leptoquark ${\cal S}_0^L$ is ruled out while
${\cal S}_1^L$ remains consistent in our global analysis.

Studies of a future scalar leptoquark search at the LHC \cite{lhc} show that
with a luminosity of 100 fb$^{-1}$ the LHC can probe leptoquark mass
up to 1.5 TeV in the pair production channel (which does not depend on the
Yukawa coupling) and up to about 3 TeV (with the Yukawa coupling the same 
as $e$) in the single production channel.  Thus, the leptoquarks in our 
solutions can be observed or ruled out at the LHC.  On the other hand, 
Run II at the Tevatron can only probe leptoquarks up to a mass of 425 GeV
\cite{teva}.

Comments about the origin of these leptoquarks are in order.  

(i) The $R$-parity violating (RPV) squarks, which arise from the supersymmetry
framework without the $R$ parity, are special leptoquarks. The natural 
question to ask is whether the leptoquarks that are used to explain the APV
measurement can be the RPV squarks.   First, since the RPV
squarks couple to leptons via the $LQD^c$ term in the superpotential, 
they only couple to the left-handed leptons.  
Therefore, ${\cal S}_{1/2}^R$, which couples to the right-handed electron,
cannot be a RPV squark.  
Also, the leptoquark ${\cal S}_1^L$, which is an isospin-triplet, is not a 
RPV squark. 
On the other hand, the leptoquark ${\cal S}_0^L$ has the interactions 
$g_L ( \overline{u_L^{(c)}} e_L - \overline{d_L^{(c)}} \nu_L )\; {\cal S}_0^L$,
which is exactly the same as the RPV squark $\tilde{d}^*_R$,
while the isospin-doublet leptoquark $\tilde{\cal S}^L_{1/2}$, which has the 
interactions $\tilde{\lambda}_L \overline{\ell_L} d_R \tilde{\cal S}^L_{1/2}$,
is equivalent to the left-handed RPV squark doublet $i \tau_2 (\tilde{u}^*_L,
\tilde{d}^*_L)^T$.  The natural question to ask is whether the coexistence 
of ${\cal S}_0^L$ and $\tilde{\cal S}_{1/2}^L$ can help 
${\cal S}_0^L$ to evade the constraint of lepton-hadron universality of weak
charged currents, and at the same time still gives a positive $\Delta Q_W$.
In Sec. 3, we have shown that ${\cal S}_0^L$ gives a 
positive $\Delta Q_W$ while $\tilde{\cal S}_{1/2}^L$ gives a negative 
$\Delta Q_W$, so that their contributions to $\Delta Q_W$ cancel.  In
fitting to the $Q_W$ measurement, the coexistence of ${\cal S}_0^L$ 
and $\tilde{\cal S}_{1/2}^L$ would give a lower ${\cal S}_0^L$ mass  or a 
higher Yukawa coupling.
However, ${\cal S}_0^L$ induces $\eta_{CC}$ as it couples to both 
left-handed leptons and quarks, while $\tilde{\cal S}_{1/2}^L$ does not because
it couples to left-handed leptons and right-handed quarks.  Therefore, the
simultaneous existence of ${\cal S}_0^L$ and $\tilde{\cal S}_{1/2}^L$ 
would not help ${\cal S}_0^L$ to evade the constraint from lepton-hadron 
universality of weak charged-currents.

(ii) 
The $F=-2$ leptoquark ${\cal S}_0^L$ is one of the leptoquarks of
$E_6$ \cite{rizzo}.
The $F=0$ leptoquark ${\cal S}^R_{1/2}$ can be embedded \cite{rizzo} 
in the flipped SU(5)$\times$U(1)$_X$ model \cite{ellis}, 
in which the SM fermion content is extended by right-handed neutrinos.
The associated right-handed neutrinos could be used to generate the 
neutrino masses by the see-saw mechanism.  
The ${\cal S}^R_{1/2}$ can be placed into ${\bf 10 + \overline{10}}$, which 
would also contain the $F=-2$ leptoquark $\tilde{\cal S}^R_0$. 
The simultaneous existence of these two leptoquarks with similar masses and
couplings would give cancelling contributions to $\Delta Q_W$.
Thus, from the view point of fitting to the $Q_W$ data, this is not favorable.

In summary, we have found leptoquark and contact interaction solutions
to the atomic parity violation measurement, which stands at a $2.3\sigma$
deviation from the SM prediction.  In addition, we have shown that these 
solutions are consistent with all other data.

\section*{\bf Acknowledgments}
This research was supported in part by the U.S.~Department of Energy under
Grants No. DE-FG03-91ER40674 and No. DE-FG02-95ER40896 and in part by the 
Davis Institute for High Energy Physics and the University of Wisconsin
Research Committee with funds granted by the Wisconsin Alumni Research
Foundation.


\end{document}